\documentclass[11pt,a4paper]{article}
\usepackage{graphicx}
\usepackage{amsmath}
\usepackage{graphicx}
\usepackage{verbatim}
\usepackage{color}
\usepackage{subfigure}
\usepackage{float}
\usepackage{authblk}
\usepackage[colorlinks]{hyperref}
\usepackage{doi}
\usepackage{textcomp}
\usepackage{eurosym}

\newtheorem{definition}{Definition}[section]

\begin{document}
\title{QSPR Analysis with Curvilinear Regression Modeling and Temperature-based Topological Indices}
\author[]{H. M. Nagesh}
\affil[]{Department of Science and Humanities, PES University, Bangalore, India. \newline E-mail: hmnagesh1982@gmail.com }

\date{}

\maketitle
\begin{abstract} Establishing quantitative correlations between various molecular properties and chemical structures is of great technological importance for environmental and medical aspects. These approaches are referred to as Quantitative Structure-Property Relationships (QSPR), which relate the physicochemical or thermodynamic properties of compounds to their structures. The main goal of QSPR studies is to find a mathematical relationship between the property of interest and several molecular descriptors derived from the structure of the molecule. Topological indices are the molecular descriptors that characterize the formation of chemical compounds and predict certain physicochemical properties. In this study, the QSPR models are designed using certain temperature-based topological indices such as the sum connectivity temperature index, product connectivity temperature index, F-temperature index, and symmetric division temperature index to predict the thermodynamic properties, such as enthalpies of formation ($\Delta H^{0}_{f}$ \hspace{1mm} liquid), enthalpies of combustion ($\Delta H^{0}_{C}$ \hspace{1mm} liquid), and enthalpies of vaporization ($\Delta H^{0}_{vap}$ \hspace{1mm} gas) of monocarboxylic acids ($C_2H_{4}O_{2}$ - $C_{20}H_{40}O_{2}$). The relationship analysis between thermodynamic properties and topological indices is done using linear, quadratic, and cubic equations of a curvilinear regression model. These regression models are then compared.     

\vskip1em \noindent \textbf{Keywords:} Temperature of a vertex, sum connectivity temperature index, product connectivity temperature index, F-temperature index, symmetric division temperature index.

\end{abstract}

\section{Introduction} \label{sec:Intr}
The properties of a molecule are indeed closely tied to its structural characteristics and composition. This concept is fundamental to understanding how molecules behave and interact in various chemical reactions and physical processes. In this connection, graph theory has been successfully applied and some thermodynamic properties \cite{1,2,6,13,16}.

Chemical graph theory is a specialized field within mathematical chemistry that focuses on the study of molecules and chemical systems through the lens of graph theory. \newpage Graph theory provides a mathematical framework for analyzing the relationships between objects connected by edges, and in the context of chemical graph theory, these objects are atoms and the edges represent chemical bonds. Chemical graph theory has practical applications in fields such as drug discovery, materials science, computational chemistry, and chemical informatics. It provides a powerful approach to understanding the relationships between molecular structure and properties, which is crucial for designing new molecules with desired characteristics and predicting how molecules will behave under different conditions. 

Chemical graph theory plays a crucial role in developing QSPR models. These models correlate graph-based molecular descriptors with various properties such as boiling points, melting points, solubility, etc. For more details, the reader can refer to \cite{3,7,9,10,11,12,15}. Numerous studies have been made relating to QSPR models by using what are called topological indices (TI). The first topological index was the Wiener index, which was introduced by Harold Wiener in 1947. It was used to determine the physical properties of paraffin \cite{16}. Since then, many topological indices have been defined and used in many applications. 

A topological index can be classified according to the structural characteristics of the graph such as the degree of a vertex, the distance between vertices, the matching, and the spectrum. The best-known topological indices are the Wiener index which is based on distance, the Zagreb and the Randi\'c indices which are based on degree, the Estrada index which is based on the spectrum of a graph, the Hosaya index based on the matching.

Shafiei in \cite{14} designed the QSPR models using topological indices such as the connectivity index, Szeged index, Balaban index, and Harary number to predict the thermodynamic properties such as enthalpies of formation of liquid, enthalpies of combustion of liquid, enthalpies of vaporization, and enthalpies of sublimation of monocarboxylic acids ($C_2H_{4}O_{2}$ - $C_{20}H_{40}O_{2}$). Later in \cite{5}, Havare designed the QSPR models using topological indices such as the Gutman index, variance of degree index,  product connectivity Banhatti index, and Sigma index to predict these thermodynamic properties of monocarboxylic acids.

Motivated by these, the structure-property relationship between certain temperature-based topological indices such as sum connectivity temperature index, product connectivity temperature index, F-temperature index, and symmetric division temperature index to the enthalpies of formation ($\Delta H^{0}_{f}$ \hspace{1mm} liquid), enthalpies of combustion ($\Delta H^{0}_{C}$ \hspace{1mm} liquid), and enthalpies of vaporization ($\Delta H^{0}_{vap}$ \hspace{1mm} gas) of monocarboxylic acids and their quantitative structure-property relationship are presented in this paper.  

\section{Basic definitions}
Let $G=(V, E)$ be a simple connected graph with vertex set $|V(G)|=n$ and edge set $E(G)$. The number of edges incident to a vertex $v$ is called the $degree$ of the vertex $v$ and is denoted by $deg(v)$. Fajtolowicz \cite{4} introduced the notion of the temperature of a vertex $v$ as follows.
\begin{definition}
\emph{The $temperature$ of a vertex $v\in V(G)$ is defined by $\mathcal{T}(v)=\frac{deg(v)}{n-deg(v)}$} 
\end{definition}
\newpage
Later in 2019, Kulli \cite{8} introduced the concept of sum connectivity temperature index, product connectivity temperature index, F-temperature index, and symmetric division temperature index of a graph $G$ as follows: \\
Sum connectivity temperature index:
\begin{equation}
S\mathcal{T}(G)=\displaystyle \sum_{uv \in E(G)}\frac{1}{\sqrt{\mathcal{T}(u)+\mathcal{T}(v)}}
\end{equation}
Product connectivity temperature index:
\begin{equation}
P\mathcal{T}(G)=\displaystyle \sum_{uv \in E(G)}\frac{1}{\sqrt{\mathcal{T}(u)\times \mathcal{T}(v)}}
\end{equation}
F-temperature index:
\begin{equation}
F\mathcal{T}(G)=\displaystyle \sum_{uv \in E(G)} 
 \left(\mathcal{T}(u)^2+\mathcal{T}(v)^2 \right)
\end{equation}
Symmetric division temperature index:
\begin{equation}
SD\mathcal{T}(G)=\displaystyle \sum_{uv \in E(G)}  \left(\frac{\mathcal{T}(u)}{\mathcal{T}(v)}+\frac{\mathcal{T}(v)}{\mathcal{T}(u)} \right)
\end{equation}

\section{Methods and techniques}
The method used in this article includes finding the temperature of vertices, division of temperature of vertices, and partition of edges based on the temperature of end vertices. The temperature of vertices and the edge separation methods are used for the computation of temperature-based topological indices. The correlation coefficients are calculated using JMP statistical software. 2D graphs are drawn using JMP statistical software/Microsoft Excel.
\section{2D molecular structures and computations of monocarboxylic acids}
With the help of these four temperature-based topological indices, the molecular structures of 19 monocarboxylic acids are explored. There are three thermodynamic properties of monocarboxylic acids under observation. These properties are enthalpies of formation ($\Delta H^{0}_{f}$ \hspace{1mm} liquid), enthalpies of combustion ($\Delta H^{0}_{C}$ \hspace{1mm} liquid), and enthalpies of vaporization ($\Delta H^{0}_{vap}$ \hspace{1mm} gas). Thermodynamic properties of monocarboxylic acids as given in Table 1 are taken from \cite{5,14}.

Let $\alpha$=Enthalpies of formation of liquids; $\beta$=Enthalpies of combustion of liquids; and $\Gamma$=Enthalpies of vaporization. Then,
\newpage
\begin{table}[h!]
\centering
\begin{tabular}{||c |c |c |c |c||} 
 \hline
 Name of compounds & Formula & $\alpha$ & $\beta$ & $\Gamma$ \\ [0.5ex] 
 \hline\hline
 Acetic acid & $C_2H_4O_2$ & 483.50  &875.16 &  46.3\\
 \hline
Propanoic acid & $C_3H_6O_2$ & 510.8 & 1527.3     & 50\\
\hline
Butanoic acid & $C_4H_8O_2$ & 533.9      & 2183.5 & 54.9\\
\hline
Pentanoic acid & $C_5H_{10}O_2$  & 558.9  & 2837.8 & 58.2\\
\hline
Hexanoic acid & $C_6H_{12}O_2$ & 581.8 & 3494.3 & 63 \\
\hline
Heptanoic acid & $C_7H_{14}O_2$ & 608.5      & 4146.9 & 64.8\\
\hline
Octanoic acid & $C_8H_{16}O_2$   & 634.8 & 4799.9   & 69.4\\
\hline
Nonanoic acid & $C_9H_{18}O_2$  & 658 &  5456.1  & 72.3\\
\hline
Decanoic acid &  $C_{10}H_{20}O_2$  & 713.7 & 6079.3    &76.6\\
\hline
Undecanoic acid &   $C_{11}H_{22}O_2$  & 736.2 &  6376.5  &78.9\\
\hline
Dodecanoic acid &  $C_{12}H_{24}O_2$  & 775.1   & 7377 & 82.2\\
\hline
Tridecanoic acid  &  $C_{13}H_{26}O_2$  & 807.2 &  8024.2    & 84.9\\
\hline
Tetradecanoic acid  &  $C_{14}H_{28}O_2$  & 834.1  &  8676.7   & 87.7\\
\hline
Pentadecanoic acid  &  $C_{15}H_{30}O_2$  & 862.4   &  9327.7    &  91.4\\
\hline
Hexadecanoic acid  &  $C_{16}H_{32}O_2$  & 892.2 &  9977.2     & 94.5\\
\hline
Heptadecanoic acid  &  $C_{17}H_{34}O_2$  & 924.4 & 10624.4    & 100.7 \\
\hline
Octadecanoic acid &  $C_{18}H_{36}O_2$  & 947.2   & 11280.1    & 102.8\\
\hline
Nonadecanoic acid  &  $C_{19}H_{38}O_2$  & 984.1 &  11923.4  & 105 \\
\hline
Eicosanoic acid &  $C_{20}H_{40}O_2$  & 1012.6  & 12574.2 & 109.9\\ [1ex] 
 \hline
\end{tabular}
\end{table} 
\textbf{Table 1}. The values of enthalpies of formation of liquid ($\Delta H^{0}_{f} kJ/mol$), enthalpies of combustion of liquid ($\Delta H^{0}_{C} kJ/mol$), and enthalpies of vaporization ($\Delta H^{0}_{vap} kJ/mol$ of monocarboxylic acids at conditions, normally at 298.15 K, 1 atm.
\section{Computation of temperature-based topological indices}
Table 2 shows the temperature-based topological indices of 19 monocarboxylic acids calculated using the formulas (1-4). 
\newpage
\begin{table}[h!]
\centering
\begin{tabular}{||c |c |c |c| c ||} 
 \hline
 Formula & $S\mathcal{T}(G)$ & $P\mathcal{T}(G)$ & $F\mathcal{T}(G)$ & $SD\mathcal{T}(G)$\\ [0.5ex] 
 \hline\hline
$C_2H_4O_2$ & 1.643167	&3	& 27.3333333	& 27.3333333  \\
\hline
$C_3H_6O_2$ & 3.23569	& 6.715476	&7.826389	& 18.069445 \\
\hline
$C_4H_8O_2$ & 4.837467	&11.0486 	& 4.12	&17.8 \\
\hline
$C_5H_{10}O_2$  & 6.585927	&16.355579	&2.730833	&18.669444 \\
\hline
$C_6H_{12}O_2$ & 8.478948	&22.649944	&2.030113	&19.993658 \\
\hline
$C_7H_{14}O_2$ & 10.5099	& 29.9373	& 1.6132 	& 21.5446\\
\hline
$C_8H_{16}O_2$   & 12.671578	&38.220202	&1.338057	&23.224868\\
\hline
$C_9H_{18}O_2$  & 14.957559	&47.500261	&1.143533	&24.985648 \\
\hline
$C_{10}H_{20}O_2$  & 17.700444	& 57.7783	  &0.998127	 &26.8 \\
\hline
$C_{11}H_{22}O_2$  & 19.879457	&69.054894	  &0.885875	 &28.651768 \\
\hline
$C_{12}H_{24}O_2$  & 22.505505	&81.330388	  &0.796447	 &30.530691\\
\hline
$C_{13}H_{26}O_2$  & 25.235916	&94.605041	  &0.723516	 &32.429945 \\
\hline
$C_{14}H_{28}O_2$  & 28.066954	&108.8790	  &0.662893	 &34.344811\\
\hline
$C_{15}H_{30}O_2$  & 30.995246	&124.1525061	  &0.611696	 &36.271925 \\
\hline
$C_{16}H_{32}O_2$  & 34.017723	&140.425548	  &0.567881	 &38.208823\\
\hline
$C_{17}H_{34}O_2$  & 37.131587	&157.698239	  &0.529953	 &40.153664\\
\hline
$C_{18}H_{36}O_2$  & 40.334265	&175.9706351	  &0.496798	 &42.105034 \\
\hline
$C_{19}H_{38}O_2$  & 43.62339	&195.2427855	  &0.467565	 &44.061842  \\
\hline
$C_{20}H_{40}O_2$  & 46.99677	&215.514726	  &0.441595	 &46.0232 \\ [1ex] 
 \hline
\end{tabular}
\end{table} 
\textbf{Table 2}. The values of temperature-based topological indices of monocarboxylic acids.
\section{Regression models}
Regression analysis is a statistical method that shows the relationship between two or more variables. 

To study the relationship between thermodynamic properties of monocarboxylic acids and the temperature-based topological indices, the following equations from \cite{14} are used. \\
Linear equation:
\begin{equation*}
Y=a+b_{1}X_1; \hspace{3mm} n, R^2, s, F
\end{equation*}
Quadratic equation:
\begin{equation*}
Y=a+b_{1}X_1+b_2X_{1}^{2}; \hspace{3mm} n, R^2, s, F
\end{equation*}
Cubic equation:
\begin{equation*}
Y=a+b_{1}X_1+b_2X_{1}^{2}+b_3X_{1}^{3}; \hspace{3mm} n, R^2, s, F
\end{equation*}

Here, $Y$ is the dependent variable, $a$ is the regression model, $b_i$ $(1 \leq i \leq 3)$ are the coefficients for the individual descriptor, $X_i$ $(1 \leq i \leq 3)$ are the independent variables, $n$ is the number of samples used for building the regression equation, $R^{2}$ is the correlation coefficient, $s$ is the standard error of deviation, and $F$ is the calculated value of the F-ration test. 
\newpage Note that the quality of a QSPR model can be conveniently measured by the correlation coefficient ($R^2$). A good QSPR model must have $R^{2}>0.99$.
The observed values and model predictions must be compared to measure the predictive quality of the model. So, we deal with the RMSE (Root Mean Square Error) metric for the predictive power of the model. The best predictive model is the minimum error, i.e. the minimum RMSE. 

Furthermore, $R^2$ and $F$ parameters will be considered for the goodness of fit of the model. The best goodness of fit in models is selected by using the parameters either max ($R^{2}$) or max ($F$). 

We now determine the linear, quadratics, and cubic curvilinear models of the sum connectivity temperature index, product connectivity temperature index, F-temperature index, and symmetric division temperature index for $\Delta H^{0}_{f}$, $\Delta H^{0}_{C}$, and $\Delta H^{0}_{vap}$. 

The significance of bold numbers in each table denote highest correlation value.

\subsection{Curvilinear regression models of $S\mathcal{T}(G)$, $P\mathcal{T}(G)$ , $F\mathcal{T}(G)$, $SD\mathcal{T}(G)$ for $\Delta H^{0}_{f}$ } 

\textbf{Table 3}. The curvilinear regressions models of $S\mathcal{T}(G)$ index for $\Delta H_{f}^{0}$.

\begin{table}[h!]
\centering
\begin{tabular}{||c |c |c |c||} 
 \hline
 $\Delta H_{f}^{0}$   & $R^{2}$ & $F$ & RMSE
 \\ [0.5ex] 
 \hline\hline
485.412+11.8001$(S\mathcal{T}(G))$ & 0.9918	&2056.329	& 15.8062 \\
\hline
456.784+15.606$(S\mathcal{T}(G))$-0.08110$(S\mathcal{T}(G)^{2})$ & \textbf{0.9988}	& 6658.513 	&6.2330	 \\
\hline
460.745+14.626$(S\mathcal{T}(G))$-0.0296$(S\mathcal{T}(G)^{2})$ -0.000717$(S\mathcal{T}(G)^{3})$ & \textbf{0.9988}	& 4449.708	& 6.2257 \\[1ex] 
 \hline
\end{tabular}
\end{table} 
 
The analysis as mentioned in Table 3 indicates that the best goodness of fit among obtained curvilinear equations using $S\mathcal{T}(G)$ topological index for $\Delta H_{f}^{0}$ are:\\
$\Delta H_{f}^{0}=456.784+15.606 (S\mathcal{T}(G))-0.08110 (S\mathcal{T}(G)^{2})$; and\\
$\Delta H_{f}^{0}=460.745+14.626 (S\mathcal{T}(G))-0.0296(S\mathcal{T}(G)^{2}) -0.000717(S\mathcal{T}(G)^{3})$.\\\\
\textbf{Table 4}. The curvilinear regressions models of $P\mathcal{T}(G)$ index for $\Delta H_{f}^{0}$.

\begin{table}[h!]
\centering
\begin{tabular}{||c| c| c |c||} 
 \hline
 $\Delta H_{f}^{0}$   & $R^{2}$ & $F$ & RMSE
 \\ [0.5ex] 
 \hline\hline
532.892+2.465$(P\mathcal{T}(G))$ & 0.9662	&487.167	& 32.053 \\
\hline
489.678+4.024$(P\mathcal{T}(G))$-0.0077$(P\mathcal{T}(G)^{2})$ & 0.9972	& 2950.84 	& 9.355	 \\
\hline
479.340+4.754$(P\mathcal{T}(G))$-0.0168$(P\mathcal{T}(G)^{2})$ -2.995e-5$(P\mathcal{T}(G)^{3})$ & \textbf{0.9986}	& 3753.734	& 6.777 \\[1ex] 
 \hline
\end{tabular}
\end{table} 

The analysis as mentioned in Table 4 indicates that the best goodness of fit among obtained curvilinear equations using $P\mathcal{T}(G)$ topological index for $\Delta H_{f}^{0}$ is:\\
$\Delta H_{f}^{0}=479.340+4.754(P\mathcal{T}(G))-0.0168(P\mathcal{T}(G)^{2}) -2.995e-5(P\mathcal{T}(G)^{3})$ \newpage
\textbf{Table 5}. The curvilinear regressions models of $F\mathcal{T}(G)$ index for $\Delta H_{f}^{0}$. 

\begin{table}[h!]
\centering
\begin{tabular}{||c |c| c| c||} 
 \hline
 $\Delta H_{f}^{0}$  & $R^{2}$ & $F$ & RMSE
 \\ [0.5ex] 
 \hline\hline
782.969-14.820$(F\mathcal{T}(G))$ & 0.2913	& 6.9901	& 146.942 \\
\hline
873.844-87.507$(F\mathcal{T}(G))$+2.692$(F\mathcal{T}(G)^{2})$ & 0.6220	& 13.1658 	& 110.619	 \\
\hline
1013.011-271.1496$(F\mathcal{T}(G))$+34.177$(F\mathcal{T}(G)^{2})$-0.9134$(F\mathcal{T}(G)^{3})$ & \textbf{0.8510}	& 28.574	& 71.713 \\[1ex] 
 \hline
\end{tabular}
\end{table} 
 
The analysis as mentioned in Table 5 indicates that the best goodness of fit among obtained curvilinear equations using $F\mathcal{T}(G)$ topological index for $\Delta H_{f}^{0}$ is:\\
$\Delta H_{f}^{0}=1013.011-271.1496(F\mathcal{T}(G))+34.177(F\mathcal{T}(G)^{2})-0.9134(F\mathcal{T}(G)^{3})$.\\\\
\textbf{Table 6}. The curvilinear regressions models of $SD\mathcal{T}(G)$ index for $\Delta H_{f}^{0}$. 
\begin{table}[h!]
\centering
\begin{tabular}{||c |c |c |c||} 
 \hline
 $\Delta H_{f}^{0}$   & $R^{2}$ & $F$ & RMSE
 \\ [0.5ex] 
 \hline\hline
224.623+17.044$(SD\mathcal{T}(G))$ & 0.8630	& 107.0981	& 64.607 \\
\hline
276.589+13.4216$(SD\mathcal{T}(G))$+0.0578$(SD\mathcal{T}(G)^{2})$ & 0.8636	& 50.6906 	& 66.429	 \\
\hline
653.886-26.121$(SD\mathcal{T}(G))$+1.3603$(SD\mathcal{T}(G)^{2})$-0.0135$(SD\mathcal{T}(G)^{3})$ & \textbf{0.8658}	& 32.275	& 68.059 \\[1ex] 
 \hline
\end{tabular}
\end{table} 

The analysis as mentioned in Table 6 indicates that the best goodness of fit among obtained curvilinear equations using $SD\mathcal{T}(G)$ topological index for $\Delta H_{f}^{0}$ is:\\
$\Delta H_{f}^{0}=653.886-26.121(SD\mathcal{T}(G))+1.3603(SD\mathcal{T}(G)^{2})-0.0135(SD\mathcal{T}(G)^{3})$.

From the analysis as mentioned in Table 3 - Table 6, the topological indices $S\mathcal{T}(G)$ and $P\mathcal{T}(G)$ are the best suitable for predicting the $\Delta H_{f}^{0}$ of monocarboxylic acids ($C_2H_{4}O_{2}$ - $C_{20}H_{40}O_{2}$) since $R^2 > 0.99$.

\subsection{Curvilinear regression models of $S\mathcal{T}(G)$, $P\mathcal{T}(G)$ , $F\mathcal{T}(G)$, $SD\mathcal{T}(G)$ for $\Delta H^{0}_{C}$ } 
\textbf{Table 7}. The curvilinear regressions models of $S\mathcal{T}(G)$ index for $\Delta H_{C}^{0}$. 

\begin{table}[h!]
\centering
\begin{tabular}{||c |c |c |c||} 
 \hline
 $\Delta H_{C}^{0}$   & $R^{2}$ & $F$ & RMSE
 \\ [0.5ex] 
 \hline\hline
1258.293+253.772$(S\mathcal{T}(G))$ & 0.9889	&1526.730	& 394.502 \\
\hline
512.470+352.946$(S\mathcal{T}(G))$-2.1129$(S\mathcal{T}(G)^{2})$ & 0.9992	& 1038.96 	&107.495	 \\
\hline
274.317+411.859$(S\mathcal{T}(G))$-5.2041$(S\mathcal{T}(G)^{2})$ +0.0431$(S\mathcal{T}(G)^{3})$ & \textbf{0.9998}	& 3037.92	& 51.341 \\[1ex] 
 \hline
\end{tabular}
\end{table} 

The analysis as mentioned in Table 7 indicates that the best goodness of fit among obtained curvilinear equations using $S\mathcal{T}(G)$ topological index for $\Delta H_{C}^{0}$ is:\\
$\Delta H_{C}^{0}=274.317+411.859(S\mathcal{T}(G))-5.2041(S\mathcal{T}(G)^{2}) +0.0431(S\mathcal{T}(G)^{3})$.\newpage
\textbf{Table 8}. The curvilinear regressions models of $P\mathcal{T}(G)$ index for $\Delta H_{C}^{0}$. 

\begin{table}[h!]
\centering
\begin{tabular}{||c |c |c |c||} 
 \hline
 $\Delta H_{C}^{0}$   & $R^{2}$ & $F$ & RMSE
 \\ [0.5ex] 
 \hline\hline
2288.883+52.909$(P\mathcal{T}(G))$ & 0.9594	& 402.091	& 757.149 \\
\hline
1304.786+88.4016$(P\mathcal{T}(G))$-0.1753$(P\mathcal{T}(G)^{2})$ & 0.9941	& 1351.173 	& 297.2921	 \\
\hline
920.142+115.588$(P\mathcal{T}(G))$-0.5165$(P\mathcal{T}(G)^{2})$ +0.0010$(P\mathcal{T}(G)^{3})$ & \textbf{0.9982}	& 2795.027	& 169.119 \\[1ex] 
 \hline
\end{tabular}
\end{table} 
 
The analysis as mentioned in Table 8 indicates that the best goodness of fit among obtained curvilinear equations using $P\mathcal{T}(G)$ topological index for $\Delta H_{C}^{0}$ is:\\
$\Delta H_{C}^{0}=1304.786+88.4016(P\mathcal{T}(G))-0.1753(P\mathcal{T}(G)^{2})$.\\\\
\textbf{Table 9}. The curvilinear regressions models of $F\mathcal{T}(G)$ index for $\Delta H_{C}^{0}$. 
\begin{center}
\begin{table}[h!]
\centering
\begin{tabular}{||c |c |c |c||} 
 \hline
 $\Delta H_{C}^{0}$  & $R^{2}$ & $F$ & RMSE
 \\ [0.5ex] 
 \hline\hline
7702.569-334.266$(F\mathcal{T}(G))$ & 0.3195	& 7.9835	& 3101.051 \\
\hline
9679.411-1915.462$(F\mathcal{T}(G))$+58.562$(F\mathcal{T}(G)^{2})$ & 0.6569	& 15.3177 	& 2269.749	 \\
\hline
12543.022-5694.203$(F\mathcal{T}(G))$+706.436$(F\mathcal{T}(G)^{2})$-18.795$(F\mathcal{T}(G)^{3})$ & \textbf{0.8659}	& 32.312	& 1465.029 \\[1ex] 
 \hline
\end{tabular}
\end{table} 
\end{center} 
The analysis as mentioned in Table 9 indicates that the best goodness of fit among obtained curvilinear equations using $F\mathcal{T}(G)$ topological index for $\Delta H_{C}^{0}$ is:\\
$\Delta H_{C}^{0}=12543.022-5694.203(F\mathcal{T}(G))+706.436(F\mathcal{T}(G)^{2})-18.795(F\mathcal{T}(G)^{3})$.\\\\
\textbf{Table 10}. The curvilinear regressions models of $SD\mathcal{T}(G)$ index for $\Delta H_{C}^{0}$. 

\begin{table}[h!]
\centering
\begin{tabular}{||c |c |c| c||} 
 \hline
 $\Delta H_{C}^{0}$   & $R^{2}$ & $F$ & RMSE
 \\ [0.5ex] 
 \hline\hline
-4241.663+362.973$(SD\mathcal{T}(G))$ & 0.8437	& 91.831	& 1485.794 \\
\hline
-3123.265+284.995$(SD\mathcal{T}(G))$+1.2453$(SD\mathcal{T}(G)^{2})$ & 0.8444	& 43.4386 	& 1528.186	 \\
\hline
1084.461-156.0012$(SD\mathcal{T}(G))$+15.7711$(SD\mathcal{T}(G)^{2})$-0.1515$(SD\mathcal{T}(G)^{3})$ & \textbf{0.8450} 	& 27.269	& 1575.346 \\[1ex] 
 \hline
\end{tabular}
\end{table} 

The analysis as mentioned in Table 10 indicates that the best goodness of fit among obtained curvilinear equations using $SD\mathcal{T}(G)$ topological index for $\Delta H_{C}^{0}$ is:\\
$\Delta H_{C}^{0}=1084.461-156.0012(SD\mathcal{T}(G))+15.7711(SD\mathcal{T}(G)^{2})-0.1515(SD\mathcal{T}(G)^{3})$.

From the analysis as mentioned in Table 7 - Table 10, the topological indices $S\mathcal{T}(G)$ and $P\mathcal{T}(G)$ are the best suitable for predicting the $\Delta H_{C}^{0}$ of monocarboxylic acids ($C_2H_{4}O_{2}$ - $C_{20}H_{40}O_{2}$) since $R^2 > 0.99$. 
\newpage

\subsection{Curvilinear regression models of $S\mathcal{T}(G)$, $P\mathcal{T}(G)$ , $F\mathcal{T}(G)$, $SD\mathcal{T}(G)$ for $\Delta H^{0}_{vap}$ } 

\textbf{Table 11}. The curvilinear regressions models of $S\mathcal{T}(G)$ index for $\Delta H_{vap}^{0}$. 
 
\begin{table}[h!]
\centering
\begin{tabular}{||c |c |c c||} 
 \hline
 $\Delta H_{vap}^{0}$   & $R^{2}$ & $F$ & RMSE
 \\ [0.5ex] 
 \hline\hline
49.7480+1.3369$(S\mathcal{T}(G))$ & 0.9835	&1015.590	& 2.5482 \\
\hline
45.531+1.897$(S\mathcal{T}(G))$-0.0119$(S\mathcal{T}(G)^{2})$ & 0.9952	& 1681.431 	&1.4087	 \\
\hline
42.932+2.540$(S\mathcal{T}(G))$-0.0456$(S\mathcal{T}(G)^{2})$ +0.000470$(S\mathcal{T}(G)^{3})$ & \textbf{0.9978}	& 2317.041	& 0.9811 \\[1ex] 
 \hline
\end{tabular}
\end{table}  

The analysis as mentioned in Table 11 indicates that the best goodness of fit among obtained curvilinear equations using $S\mathcal{T}(G)$ topological index for $\Delta H_{vap}^{0}$ is:\\
$\Delta H_{vap}^{0}=42.932+2.540(S\mathcal{T}(G))-0.0456(S\mathcal{T}(G)^{2}) +0.000470(S\mathcal{T}(G)^{3})$.\\\\
\textbf{Table 12}. The curvilinear regressions models of $P\mathcal{T}(G)$ index for $\Delta H_{vap}^{0}$. 

\begin{table}[h!]
\centering
\begin{tabular}{||c |c |c |c||} 
 \hline
 $\Delta H_{vap}^{0}$   & $R^{2}$ & $F$ & RMSE
 \\ [0.5ex] 
 \hline\hline
55.201+0.278$(P\mathcal{T}(G))$ & 0.9522	& 338.806	& 4.341 \\
\hline
49.973+0.4670$(P\mathcal{T}(G))$-0.00039$(P\mathcal{T}(G)^{2})$ & 0.9872	& 621.4481 	& 2.3079	 \\
\hline
47.306+0.655$(P\mathcal{T}(G))$-0.0032$(P\mathcal{T}(G)^{2})$ +7.510$(P\mathcal{T}(G)^{3})$ & \textbf{0.9943}	& 880.0532	& 1.5891 \\[1ex] 
 \hline
\end{tabular}
\end{table} 

The analysis as mentioned in Table 12 indicates that the best goodness of fit among obtained curvilinear equations using $P\mathcal{T}(G)$ topological index for $\Delta H_{vap}^{0}$ is:\\
$\Delta H_{vap}^{0}=47.306+0.655(P\mathcal{T}(G))-0.0032(P\mathcal{T}(G)^{2}) +7.510(P\mathcal{T}(G)^{3})$.\\\\
\textbf{Table 13}. The curvilinear regressions models of $F\mathcal{T}(G)$ index for $\Delta H_{vap}^{0}$. 

\begin{center} 
\begin{table}[h!]
\centering
\begin{tabular}{||c |c |c| c||} 
 \hline
 $\Delta H_{vap}^{0}$  & $R^{2}$ & $F$ & RMSE
 \\ [0.5ex] 
 \hline\hline
83.876-1.822$(F\mathcal{T}(G))$ & 0.3402	& 8.7681	& 16.1314 \\
\hline
94.306-10.164$(F\mathcal{T}(G))$+0.308$(F\mathcal{T}(G)^{2})$ & 0.6767	& 16.7478 	& 11.6393	 \\
\hline
108.734-29.203$(F\mathcal{T}(G))$+3.5733$(F\mathcal{T}(G)^{2})$-0.09470$(F\mathcal{T}(G)^{3})$ & \textbf{0.8669}	& 32.572	& 7.7128 \\[1ex] 
 \hline
\end{tabular}
\end{table} 
\end{center}
The analysis as mentioned in Table 13 indicates that the best goodness of fit among obtained curvilinear equations using $F\mathcal{T}(G)$ topological index for $\Delta H_{vap}^{0}$ is:\\
$\Delta H_{vap}^{0}=108.734-29.203(F\mathcal{T}(G))+3.5733(F\mathcal{T}(G)^{2})-0.09470(F\mathcal{T}(G)^{3})$.\newpage
\textbf{Table 14}. The curvilinear regressions models of $SD\mathcal{T}(G)$ index for $\Delta H_{vap}^{0}$.
\begin{table}[h!]
\centering
\begin{tabular}{||c |c |c| c||} 
 \hline
 $\Delta H_{vap}^{0}$   & $R^{2}$ & $F$ & RMSE
 \\ [0.5ex] 
 \hline\hline
21.241+1.896$(SD\mathcal{T}(G))$ & 0.8255	& 80.4709	& 8.2942 \\
\hline
28.6641+1.3792$(SD\mathcal{T}(G))$+0.00826$(SD\mathcal{T}(G)^{2})$ & 0.8266 & 38.1525 	& 8.5231	 \\
\hline
36.2095+0.5884$(SD\mathcal{T}(G))$+0.03431$(SD\mathcal{T}(G)^{2})$-0.08027$(SD\mathcal{T}(G)^{3})$ & \textbf{0.8267}	& 23.8565	& 80800973 \\[1ex] 
 \hline
\end{tabular}
\end{table} 

The analysis as mentioned in Table 14 indicates that the best goodness of fit among obtained curvilinear equations using $SD\mathcal{T}(G)$ topological index for $\Delta H_{vap}^{0}$ is:\\
$\Delta H_{vap}^{0}=36.2095+0.5884(SD\mathcal{T}(G))+0.03431(SD\mathcal{T}(G)^{2})-0.08027(SD\mathcal{T}(G)^{3})$.

From the analysis as mentioned in Table 11 - Table 14, the topological indices $S\mathcal{T}(G)$ and $P\mathcal{T}(G)$ are the best suitable for predicting the $\Delta H_{vap}^{0}$ of monocarboxylic acids ($C_2H_{4}O_{2}$ - $C_{20}H_{40}O_{2}$) since $R^2 > 0.99$.
\section{Plots of the cubic regression equation}
The analysis as mentioned in Table 3 - Table 14 indicates that the cubic equation gives the best goodness of fit among three regression equations.
 
Figure 1 - Figure 3 shows the correlation of $S\mathcal{T}(G)$ and $P\mathcal{T}(G)$ with $\Delta H^{0}_{f}$, $\Delta H^{0}_{C}$, and $\Delta H^{0}_{vap}$ using cubic regression equation.  
\vspace{5mm}
\begin{figure}[h!]
\centering
\includegraphics[width=170mm]{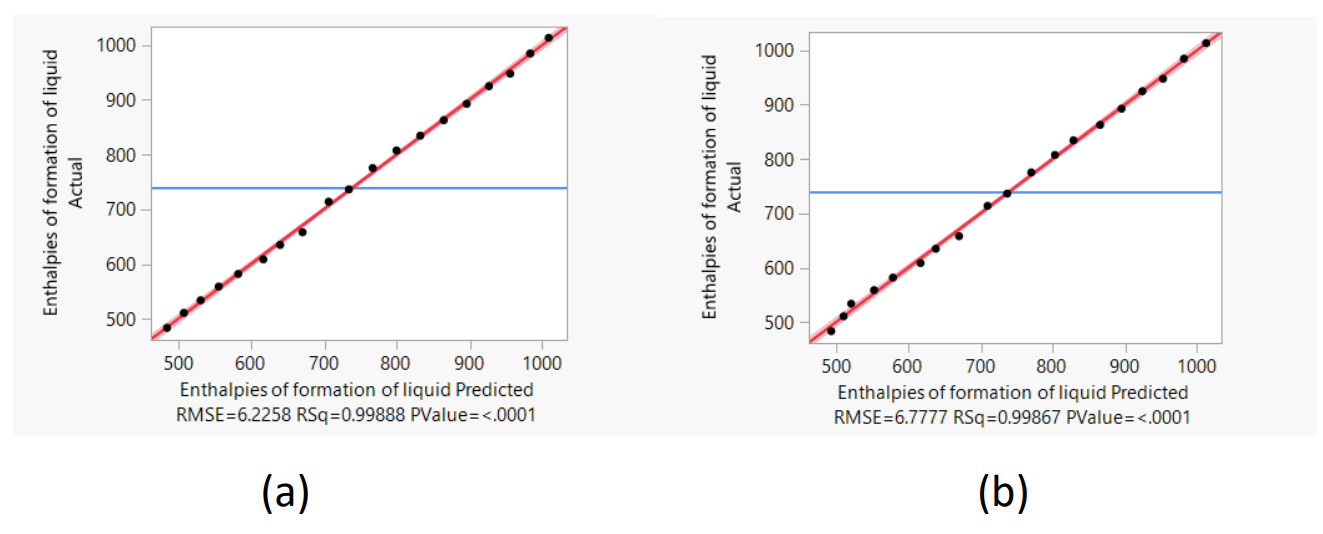}
  \caption{(a) Cubic regression equation of $\Delta H_{f}^{0}$ with $S\mathcal{T}(G)$. (b) Cubic regression equation of $\Delta H_{f}^{0}$ with $P\mathcal{T}(G)$. }
  \end{figure}  
  \newpage

\begin{figure}[h!]
\centering
\includegraphics[width=170mm]{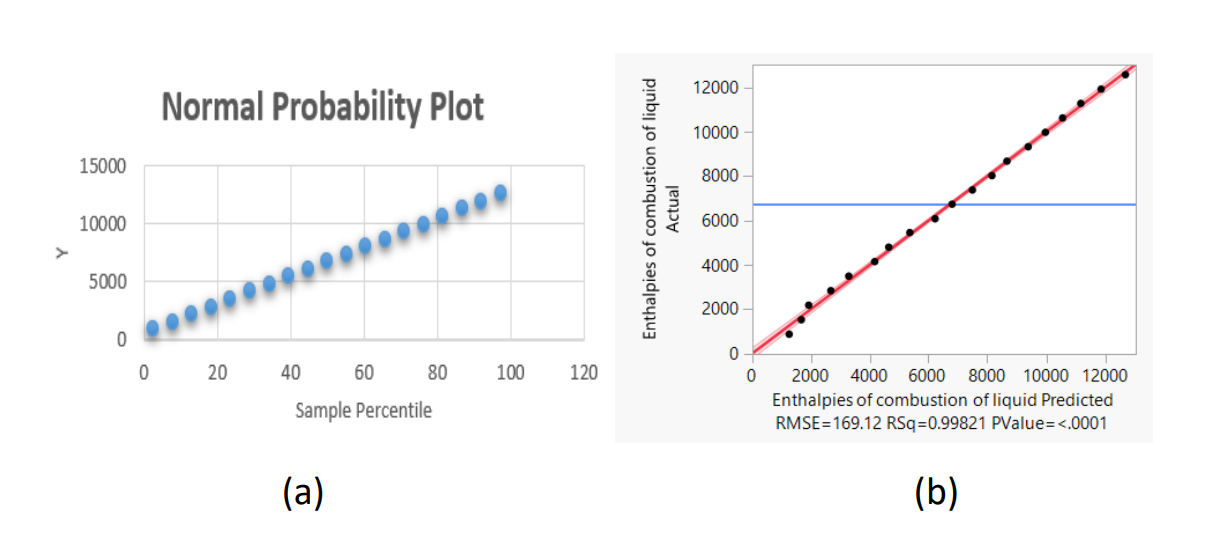}
  \caption{(a) Cubic regression equation of $\Delta H_{C}^{0}$ with $S\mathcal{T}(G)$ using normal probability plot. (b) Cubic regression equation of $\Delta H_{C}^{0}$ with $P\mathcal{T}(G)$. }
  \end{figure}

\begin{figure}[h!]
\centering
\includegraphics[width=170mm]{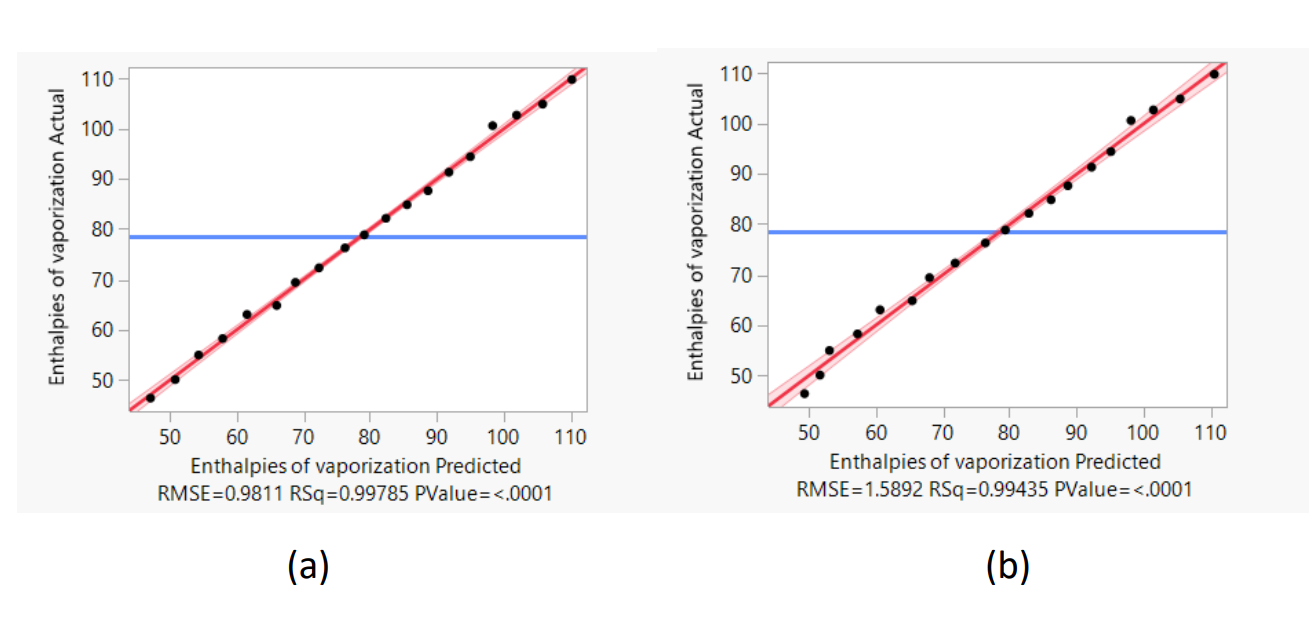}
  \caption{(a) Cubic regression equation of $\Delta H_{vap}^{0}$ with $S\mathcal{T}(G)$. (b) Cubic regression equation of $\Delta H_{vap}^{0}$ with $P\mathcal{T}(G)$. }
  \end{figure}  
\section{Conclusion}
The temperature-based topological indices such as the sum connectivity temperature index, product connectivity temperature index, F-temperature index, and symmetric division temperature index of 19 monocarboxylic acids ($C_2H_{4}O_{2}$ - $C_{20}H_{40}O_{2}$) are calculated. Using these topological indices, curvilinear regression models are designed to predict certain thermodynamic properties such as enthalpies of formation ($\Delta H^{0}_{f}$ \hspace{1mm} liquid), enthalpies of combustion ($\Delta H^{0}_{c}$ \hspace{1mm} liquid), and enthalpies of vaporization ($\Delta H^{0}_{vap}$ \hspace{1mm} gas) of monocarboxylic acids. The most accurate results for the prediction of these thermodynamic properties can be calculated by using the sum connectivity temperature index and product connectivity temperature index. Furthermore, these thermodynamic properties also have a good correlation with the symmetric division temperature index, but the F-temperature index is not enough to make a good prediction of thermodynamic properties of monocarboxylic acids ($C_2H_{4}O_{2}$ - $C_{20}H_{40}O_{2}$). The optimum is the cubic equation form among 
curvilinear equations.
\makeatletter
\renewcommand{\@biblabel}[1]{[#1]\hfill}
\makeatother

\end{document}